\documentclass{aa}
\usepackage{times}
\usepackage{graphics}

\newcommand{\bv}{\mbox{$B-V$}}
\newcommand{\bvo}{\mbox{$(B-V)_o$}}
\newcommand{\feh}{\mbox{[Fe/H]}}
\newcommand{\Msun}{\mbox{$M_{\odot}$}}
\newcommand{\Mhef}{\mbox{$M_{\rm Hef}$}}
\newcommand{\comment}[1]{}

\begin{document}

	\thesaurus{10.15.1 08.12.1 08.05.3 08.09.3 08.08.1}

        \title{On the peculiar red clump morphology in the
open clusters NGC~752 and NGC~7789 }

	\author{L\'eo Girardi$^{1}$, Jean-Claude Mermilliod$^2$,
Giovanni Carraro$^1$}
	\institute{
$^1$ Dipartimento di Astronomia, Universit\`a di Padova,
Vicolo dell'Osservatorio 5, I-35122
        Padova, Italy \\
$^2$ Institut d'Astronomie de l'Universit\'e de Lausanne, 
        CH-1290 Chavannes-des-Bois, Switzerland
	}

	\offprints{L\'eo Girardi \\ 
e-mail: Lgirardi@pd.astro.it }

	\date{Received September 1999 / Accepted January 2000}

	\maketitle
	\markboth{Girardi, Mermilliod \& Carraro}{Peculiar red giant
clump morphology}

	\begin{abstract}
The red clump stars in the open cluster NGC~752 present a peculiar 
distribution in the colour-magnitude diagram (CMD): the clump is
observed to present a faint extension, slightly to the blue of the 
main concentration of clump stars. We point out that a similar structure
is present in the CMD of NGC~7789, and discuss their possible origins.
This feature may be understood as the result of having, at the same time,
stars of low-mass which undergo the helium-flash, and those just massive
enough for avoiding it. The ages of both clusters are compatible with 
this interpretation. 

Similar features can be produced in theoretical models which assume a 
non-negligible mass spread for clump stars, of about 0.2 \Msun. However, 
one can probably exclude that the observed effect is due to the natural 
mass range of core helium burning stars found in single isochrones, 
although present models do not present the level of detail necessary to 
completely explore this possibility. Also the possibility of a large age 
spread among cluster stars can be refuted on observational grounds. 

We then suggest a few alternatives. This spread may be 
resulting either from star-to-star variations in the mass-loss rates 
during the RGB phase. Alternatively, effects such as stellar rotation or 
convective core overshooting, could be causing a significant spread in 
the core mass at He-ignition for stars of similar mass. Finally, we 
point out that similar effects could also help to understand the 
distribution of clump stars in the CMDs of the clusters NGC~2660 and 
NGC~2204.

\keywords{cluster: open - stars: evolution -- stars: interiors -- stars:
Hertzsprung--Russel (HR) diagram}

\end{abstract}  

\section{ Introduction }

The clump of red giants is a remarkable feature in the colour-magnitude
diagrams (CMD) of intermediate-age and old open clusters (Cannon~\cite{rdc}). 
It is defined by stars in the stage of core helium burning (CHeB). 

In the clusters for which the non-member 
field stars and binaries have been identified and excluded, 
the red clump may occupy a very small region of the 
CMD. A good example of this case is given by the 4-Gyr old cluster
M~67: its 6 clump stars differ in colour by less than 
0.01~mag in \bv, and 0.1~mag in $V$ (see e.g.\ Montgomery et
al.~\cite{mmj}). This small spread of the clump
can be easily understood as the result of having low-mass core
He-burning stars of very similar masses in this cluster. 

However, some clusters clearly present a more complex clump structure.
One of the best examples is given by NGC~752: Mermilliod et al.~(\cite{mmlm})
recently pointed out it presents a kind of dual clump. 
It is shown in detail in the figure 5 of Mermilliod et al.~(\cite{mmlm}):
there we can notice the presence of the main clump
centered at $\bv=1.01$, $V=9.0$ and composed of 8 member stars, and a 
distribution of  3 or 4 fainter stars, going down by about 0.5~mag in
relation with this main clump. Importantly, all
stars plotted are members of the cluster with probability $P>93$\%, 
and photometric errors are lower than 0.015 on $V$ and 0.013 on the 
colours. Therefore, the structure seen in NGC~752
is real, and not an artefact of observational uncertainties.

On the other hand, recent works suggest that clumps with a faint
extension may be a commom feature in the field of nearby galaxies. 
In a few words, population synthesis models
of galaxy fields predict that a secondary red clump may be formed at
about $0.3-0.4$~mag below the main one, containing the CHeB stars which 
are just massive enough for starting to burn helium in non-degenerate 
conditions.
Girardi et al.~\cite{ggws} first suggested the presence of this feature
in the CMD derived from {\em Hipparcos} data-base (Perryman et al.\ 
\cite{macp}; ESA \cite{esa}).
The subject has been later extensively discussed by Girardi (\cite{lg99}). 
Bica et al.\ (\cite{bgdc}) and Piatti et al.\ 
(\cite{pgbc}) recently presented clear evidence showing that this feature 
is present in the LMC.

Could this feature provide an explanation also for the peculiar 
morphology of the clump in NGC~752? Is a similar clump morphology
observed in other clusters as well? These are the main
questions we address in this paper. 

\section{The clusters}
\label{sec_clus}

First of all, we examined the available data for galactic open 
clusters, in order to identify if other clusters have clumps 
with the same general appearance as in NGC~752. 
We used the extended database of accurate photometry and radial
velocity data compiled by 
Mermilliod and collaborators. This data allows us to select the member 
stars and avoid the binaries in each cluster. In this way, clean 
CMDs can be produced. Indeed, Mermilliod \& Mayor (\cite{mm89}, \cite{mm90}) 
already noticed the presence of a number of 
stars below the clump in some clusters. The case of NGC~752,
however, is remarkable for the high fraction of stars
located below the ``main clump'' level, which causes its
apparent bi-modality.

We searched for clusters with ages similar to that of NGC~752 
($9.1 < \log t < 9.3$). Several candidates were found. 
NGC~3680 and IC~4651 (Mermilliod et al.~\cite{manm}) do not present
the same characteristics as NGC~752 does. The red giant clump in 
NGC~3680 is rather concentrated, with few scatter in magnitude and 
colours, while the morphology of IC~4651 clump is more complex. 
NGC~2158 is a rich and very interesting cluster probably showing 
also a complex structure of the clump region.
However, most data are photographic and there is presently no 
kinematical membership determination to identify the true members. 
A CCD study paying attention to the red giants would be worth.
We shall therefore refrain to use this cluster. The fourth cluster 
is NGC~7789 for which $BV$ CCD data have been published by Martinez 
Roger et al.\ (\cite{mrpc}) and Jahn et al.\ (\cite{jkr}). Gim et al.\ 
(\cite{ghmc}) have 
published an extensive radial-velocity study which permits to reject 
the non-member stars and identify the binaries. Membership 
probabilities from proper motions for NGC~7789 have been published by 
McNamara \& Solomon (\cite{mcns}) so that the membership of the red giants 
is rather well defined. Most other clusters containing numerous red 
giants and for which good photometric data are available are
either younger (about 1~Myr) and have a large clump, with a few stars
below or are older and have a more or less well developed giant
branch.

\begin{figure*}
\resizebox{\hsize}{!}{\includegraphics{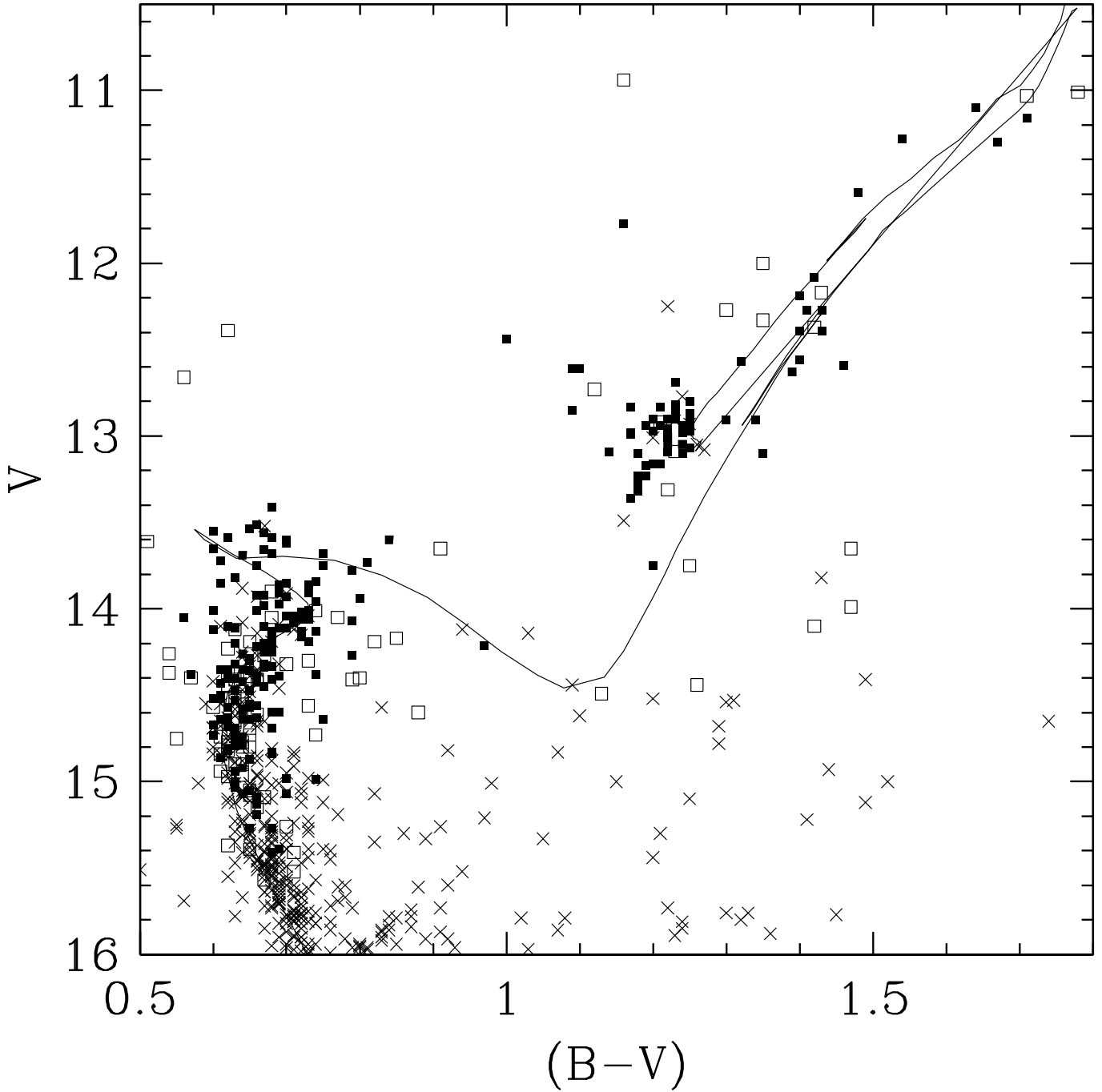}}
\caption{ 
CMD of NGC~7789. Filled squares denotes stars with membership probability
$P > 75$\%, open squares stars with $P < 75$\%, and crosses stars without 
membership probabilities. The isochrone is from the models of Girardi et al.\ 
(\cite{gbbc}) for solar metallicity and $\log t = 9.20$. Notice the fainter 
and somewhat bluer tail below the clump.}
\label{fig_ngc7789}
\end{figure*}

As can be seen in Fig.~\ref{fig_ngc7789}, 
the red clump in NGC~7789 presents a tail
of faint stars extending down to 0.4~mag below the main
concentration of clump stars. Again, the fainter clump stars are 
observed to be slightly bluer than the main clump concentration.

A comparison between NGC~752 and NGC~7789 clearly evidences
that both clusters have similar ages: suffice it to notice that 
the main sequence termination (TAMS) and red clump are observed at 
the same colour ($\bvo=0.5$ and $\bvo=1.0$, respectively), 
and that their magnitude difference is also very similar
(of about $\delta V=0.5$~mag) in both clusters. 
Also, we recall that both clusters have metallicities 
very close to each other: according to Friel \& Janes (\cite{fj83}),
$\feh=-0.16\pm0.05$ for NGC~752 and $\feh=-0.26\pm0.06$ for NGC~7789.

In Table~\ref{tab_ages}, we list a limited number of age estimates
for both clusters, in which the age-dating was based in models with 
overshooting.

\begin{table}
\caption{Age estimates for NGC~752 and NGC~7789.}
\label{tab_ages}
\begin{tabular}{lcll}
\noalign{\smallskip}\hline\noalign{\smallskip}
NGC & Age  & Method & Reference \\
    & [Gyr] & & \\
\noalign{\smallskip}\hline\noalign{\smallskip}
752
& 1.8   &      isochrone fitting &  Meynet et al.\ (\cite{mmm}) \\
& 1.5   &      synthetic CMD     &  Carraro \& Chiosi (\cite{cc94}) \\
& 2.0   &      isochrone fitting &  Dinescu et al.\ (\cite{ddgp}) \\
\noalign{\smallskip}\hline\noalign{\smallskip}
7789
&  1.3   &    synthetic  CMD      &      Carraro \& Chiosi (\cite{cc94})  \\
&  1.6   &    isochrone fitting   &      Gim et al.\ (\cite{gvsh}) \\
&  1.4   &    isochrone fitting   &      Vallenari et al.\ (\cite{vcr}) \\
\noalign{\smallskip}\hline\noalign{\smallskip}
\end{tabular}
\end{table}

It turns out that both NGC~752 and NGC7789 should have an age of 
about 1.5~Gyr. These are indicative ages, which will 
be useful in the analysis of the following sections.

\section{The models}
\label{sec_input}

\begin{figure*}
\resizebox{12cm}{!}{\includegraphics{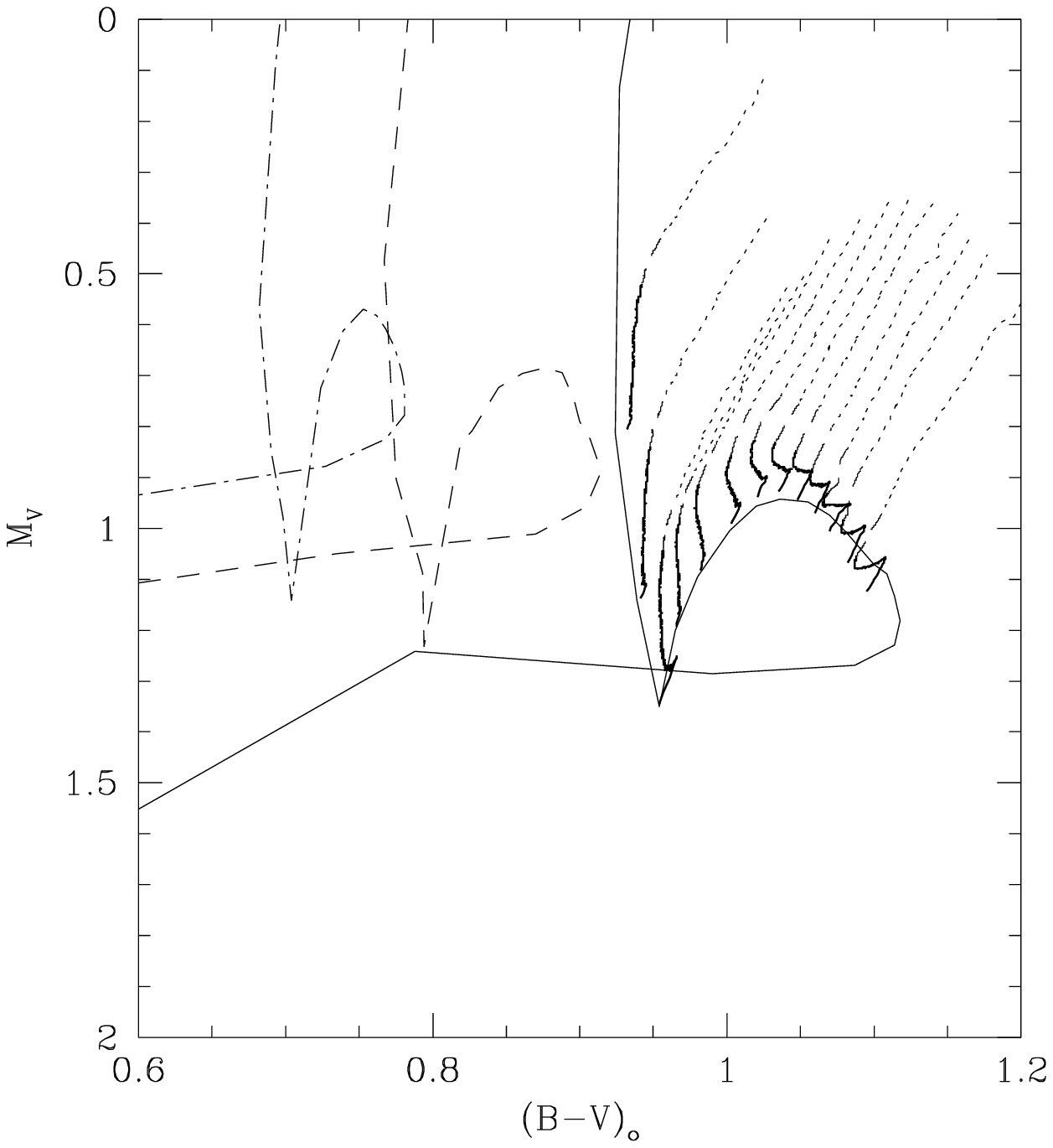}}
\hfill
\parbox[b]{55mm}{
\caption{The location of CHeB star models in the $M_V$ versus 
\bv\ diagram. In the right part of the diagram, we have a large sample 
of tracks of solar metallicity ($Z=0.019$): the continuous lines denote 
the initial 70\% 
fraction of the CHeB lifetime, for stars of different masses, whereas
the remaining 30\% fraction are marked with dashed lines. The mass range
goes from 2.5 to 1.1~\Msun\ (from left to right). The continuous line
denotes the lower boundary of the CHeB tracks as a function of initial 
mass. Similar curves denote the same lower boundary for tracks of 
metallicity $Z=0.008$ (long-dashed line) and $Z=0.004$ (dot-dashed line).
}
\label{fig_zahb}
}
\end{figure*}

In order to describe the evolution of clump stars as a function of 
cluster parameters, we make use of the stellar evolutionary tracks 
and isochrones from Girardi et al.\ (\cite{gbbc}). This data-base of stellar 
models contains CHeB stars computed for a large number of initial 
masses, providing a detailed description of the position of these stars
in both HR and colour-magnitude diagrams. Figure~\ref{fig_zahb} shows the 
location of the models for solar metallicity ($Z=0.019$) in the $M_V$ 
versus \bv\ diagram. It can be noticed that CHeB models
of varying mass distribute along a well defined sequence in this plot.
This sequence is relatively narrow if we consider the initial fraction
of 70\% of the CHeB lifetime, where most of the CHeB stars are expected 
to be found. The remaining 30\% fraction, instead, occupies a broad
distribution in the diagram. Importantly, the sequence of CHeB models
presents a well-defined lower boundary, which is also drawn in the plot. 
The same boundary line is shown for lower values of metallicity 
($Z=0.008$ and $Z=0.004$), thus showing how 
the sequences of CHeB models get bluer as the metallicity decreases.

\begin{figure*}
\resizebox{\hsize}{!}{\includegraphics{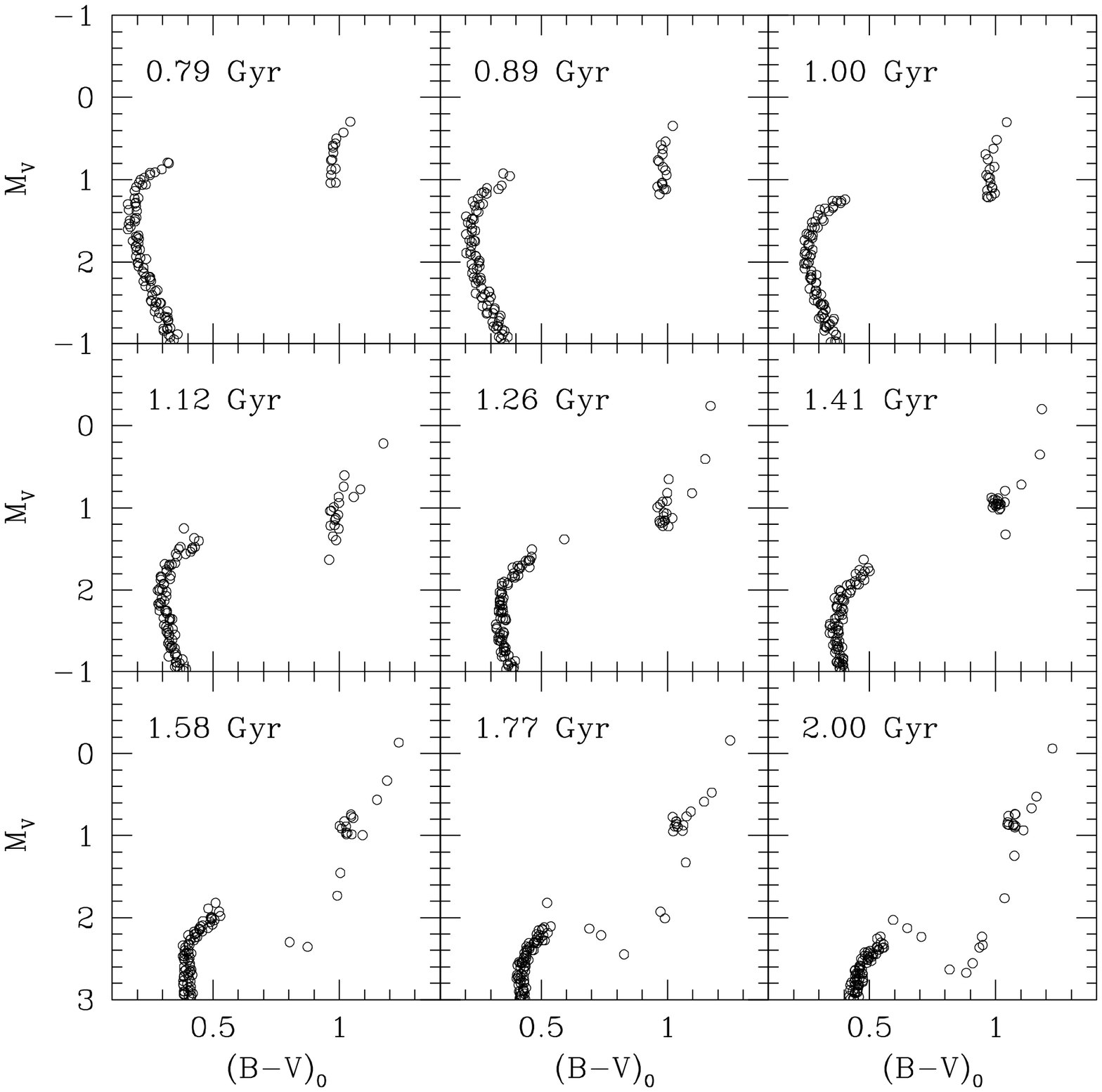}}
\caption{
Synthetic CMDs for a sequence of cluster models with solar
metallicity. Cluster ages increase by a factor of 
12 per cent from panel to panel. Notice the initial gradual 
fading of the clump, which gets again brighter between 1.26 and
1.58~Gyr.}
\label{fig_models}
\end{figure*}

When clusters become older, we expect to find
CHeB stars of lower and lower masses. Therefore, the position of 
clump stars in an ageing cluster roughly follows the sequence 
shown in Fig.~\ref{fig_zahb}, going from the upper left to the bottom 
right of the diagram. Along this sequence, however, the clump 
luminosity passes through a temporary minimum when the turn-off mass 
is of about 2~\Msun.
This effect is illustrated in Fig.~\ref{fig_models}, in which we 
simulate clusters with $Z=0.019$ (i.e.\ solar metallicity) and 
ages between 1 and 2~Gyr. This age interval
encompasses the probable ages of NGC~752 and NGC~7789.

In the sequence of simulations, the clump of 
CHeB stars decreases in luminosity as the cluster ages, up to about
1.26~Gyr. Then, up to an age of 1.56~Gyr, this luminosity increases
by as much as 0.4~mag, remaining almost constant afterwards. 
This increase in luminosity in a relatively short timescale 
corresponds to the age (and stellar initial mass)
at which the CHeB switches from quiescent ignition, to a mildly explosive
ignition (the He-flash) inside an electron degenerate core. 
The increase in luminosity is mostly due to the large increase in the 
core mass required to 
ignite helium in a degenerate core. The basic theory of this transition
is detailed in the classical work by Sweigart et al.\ (\cite{sgr}). 

Girardi et al.\ (\cite{ggws}) and Girardi (\cite{lg99}) already explored the 
consequences of this transition in the CMDs of galactic fields. 
The most impressive effect they found is that 
the stars with age of $\sim1$~Gyr may define a ``secondary clump'' feature
extending below the main clump of red giants. Of course, the suggestion that
the same feature may be present in star clusters like NGC~752 and NGC~7789
is immediate. 

In fact, the age at which the transition occurs in the models
is clearly in agreement with what is observed in the clusters: at 1.5 Gyr,
main sequence and red clump are observed at $\bvo=0.5$ and $\bvo=1.0$, 
respectively, and their magnitude difference is of 
$\delta V=0.5$~mag. These numbers are undistiguishable from those 
observed in the CMDs of NGC~752 and NGC~7789. 

However, it is also clear that the models of single-age, single-metallicity
stellar populations shown in Fig.~\ref{fig_models} do not produce 
``dual clumps'', neither clumps with fainter tails, as in the case
of galaxy models. What clump models indicate is that the clump has an 
intrinsic elongated structure for ages lower than 1.2~Gyr, getting
more concentrated at later ages, when the clump gets brighter.
Therefore, they do not provide an obvious explanation for the clump
morphologies observed in NGC~7789 and NGC~752.

\section{Observed red giant clumps}
\label{sec_facts}

\begin{figure*}
\begin{minipage}{0.48\textwidth} 
\resizebox{8.5cm}{!}{\includegraphics{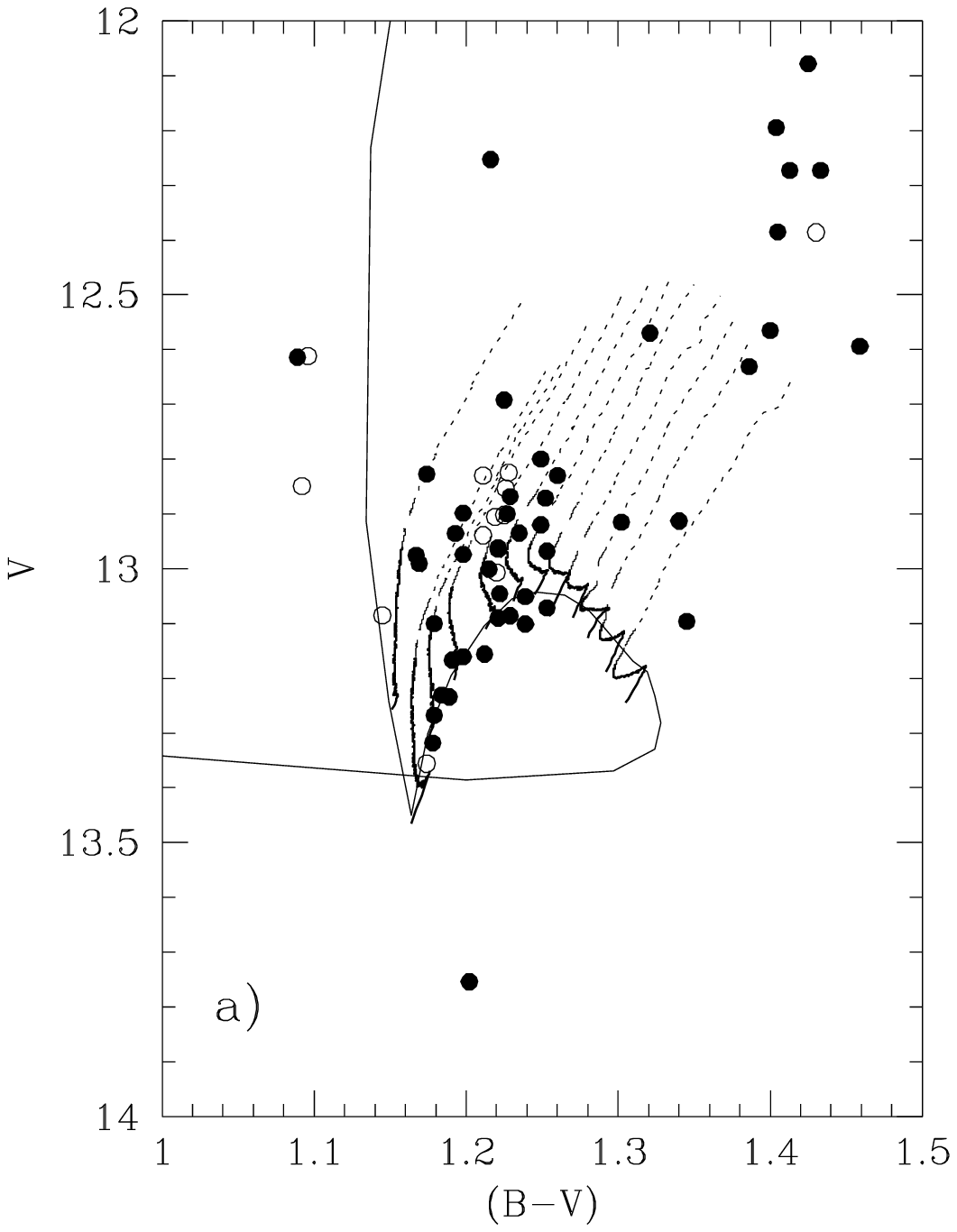}}
\end{minipage} 
\hfill
\begin{minipage}{0.48\textwidth} 
\resizebox{8.5cm}{!}{\includegraphics{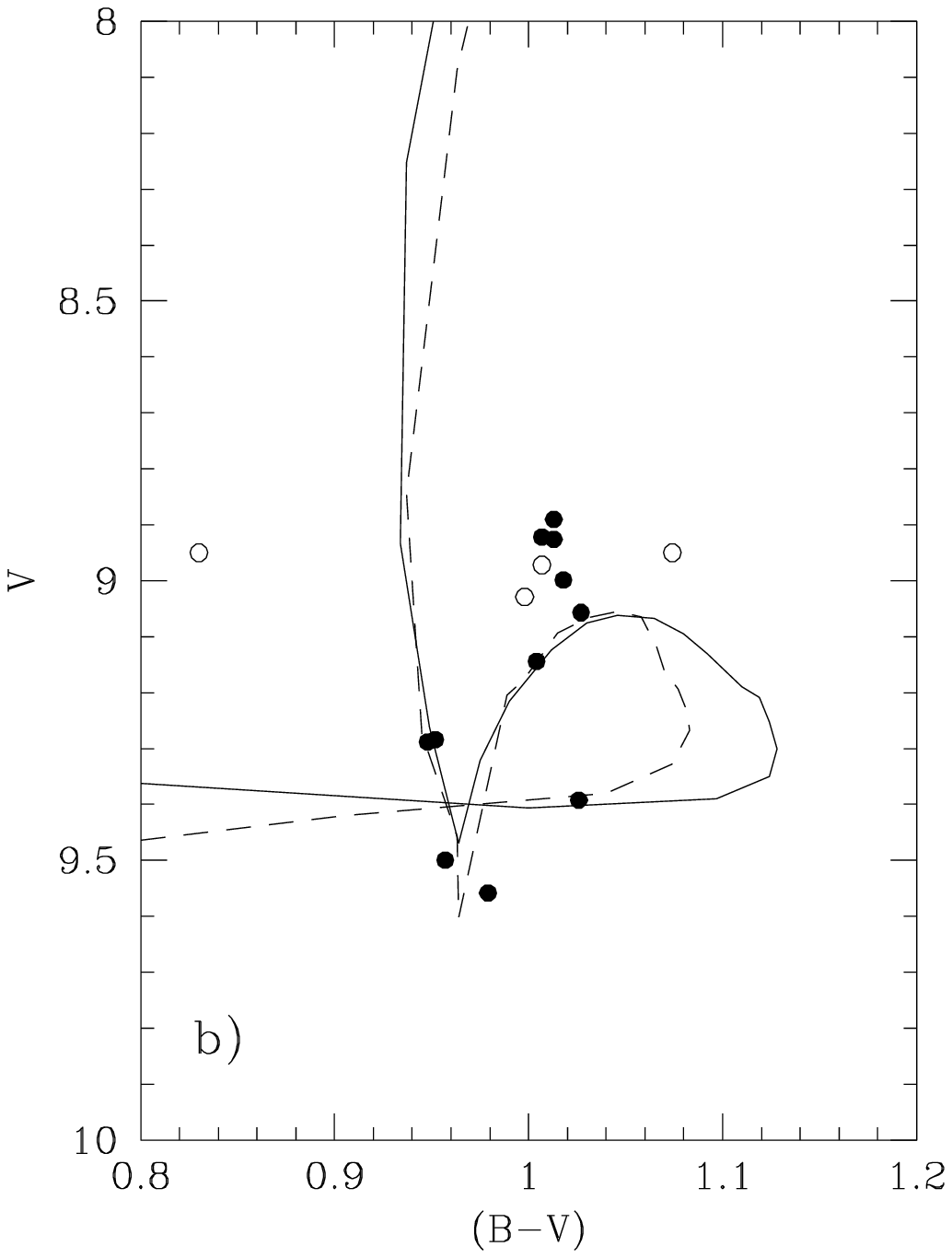}}
\end{minipage} 
\caption{ 
A small region of the CMD, centered on the clump, is shown 
for the clusters NGC~7789 (a) and NGC~752 (b).
Only probable members are plotted. Single stars are 
shown as full dots, known binaries as open symbols.}
\label{fig_clusters}
\end{figure*}

The theoretical ZAHB and individual evolutionary tracks have been 
plotted over the observed red giants in NGC~7789 
(Fig.~\ref{fig_clusters}a). Due to the shape 
of the red giant locus, the position of the ZAHB is rather obvious. 
The diagram can be interpreted as follows: a number of stars, with 
masses between 1.95 and 1.75~\Msun\ are close to or on the ZAHB, 
while other stars have already further evolved from the ZAHB. The 
bulk of the red giants has masses between 1.7 and 1.8~\Msun. 
Open circles are known binaries. Some do show the effects of the 
secondaries because their colours are bluer, while several ones are 
right in the middle of the ``clump''. The theoretical shape of the 
ZAHB does give a very good representation of the observed morphology 
of the red giant clump.

The resulting image is that at the core Helium-burning
phase we have a spread in masses and ages among the red giants.
It is evident that stars do not arrive at the same time in the
He-burning phase, but clearly the observed morphology is not 
compatible with the "classical" paradigm of evolution of single 
star, single mass isochrones. We observe, as is well known from the
evolution on the main sequence, stars on the ZAHB and stars 
leaving this phase toward the asymptiotic giant branch. What is
surprising is the size of the observed dispersion in mass on the
ZAHB.

The individual evolutionary tracks permit to understand the
vertical dispersion and assign it to the evolution away from the
ZAHB. However, the solid part of the individual evolutionary tracks, 
covering 70\% of the core Helium-burning lifetime is mostly limited 
to the very beginning of the tracks and seems to be a little too 
short with respect to the observed distribution of the stars.

The stars with $12.1< V < 12.5$ and $B-V \sim 1.4$ form a bump 
at the exact position predicted by the models (see 
Fig.~\ref{fig_ngc7789}a). It 
correponds to the phase when the H-burning shell moving outward 
encounters the H-discontinuity resulting form the first dredge-up. 
They are therefore not in an advanced core He-burning stage, but 
mark a pause in the ascent of the red giant branch.

The case of NGC~752 (Fig.~\ref{fig_clusters}b) can be understood in the 
same context. Even
if there are less red giants, the explanation seems to be quite
convincing. The ``classical'' clump is well marked, and the fainter
stars, which define the so-called second clump, are pretty 
well aligned along the ZAHB. Two models have been plotted, that
for solar metallicity ($Z=0.019$) and half solar ($Z=0.008$). If
the track is fitted to the base of the clump, both curves equally
well reproduce the positions of the points. 

\begin{figure*}
\begin{minipage}{0.48\textwidth} 
\resizebox{8.5cm}{!}{\includegraphics{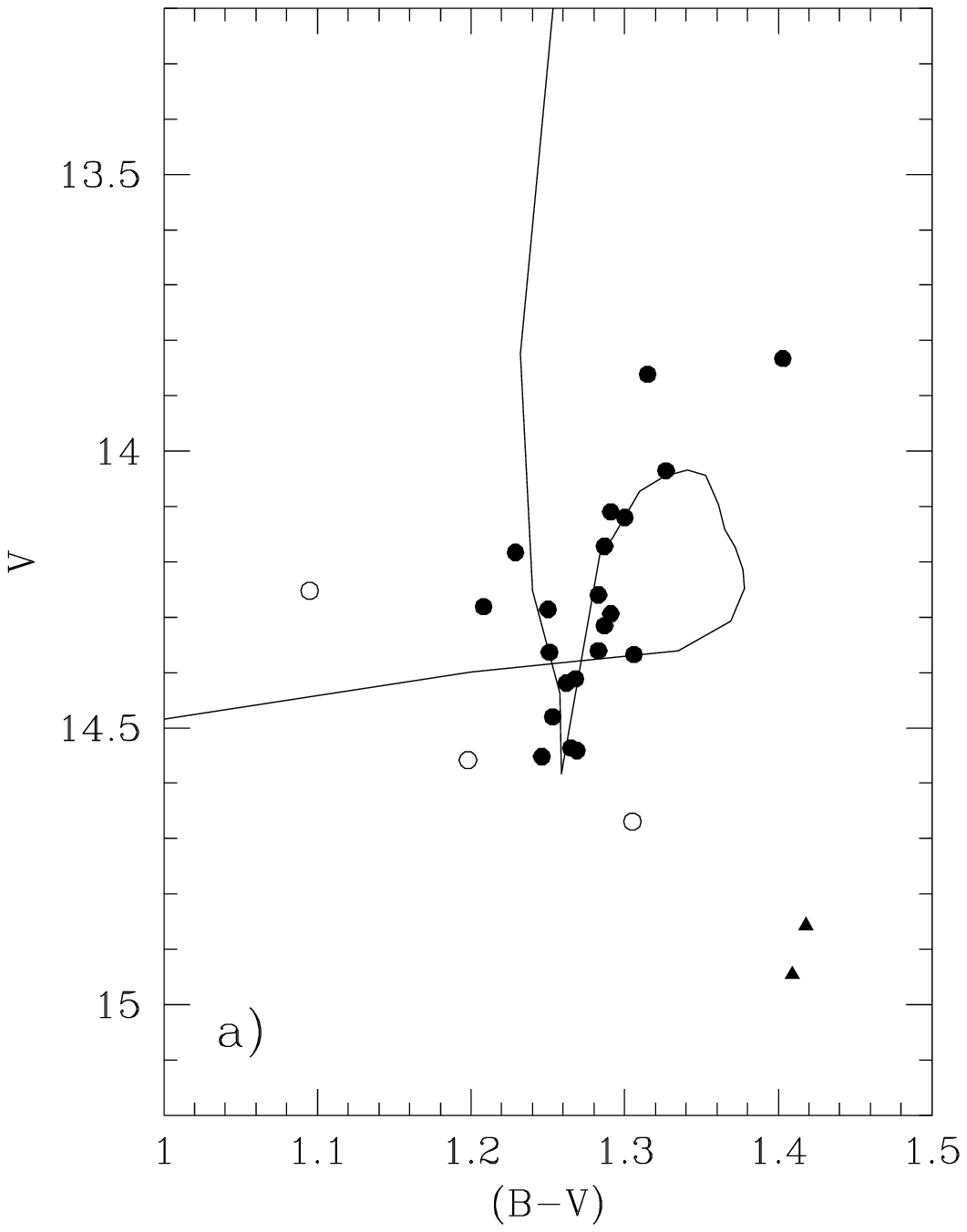}}
\end{minipage} 
\hfill
\begin{minipage}{0.48\textwidth} 
\resizebox{8.5cm}{!}{\includegraphics{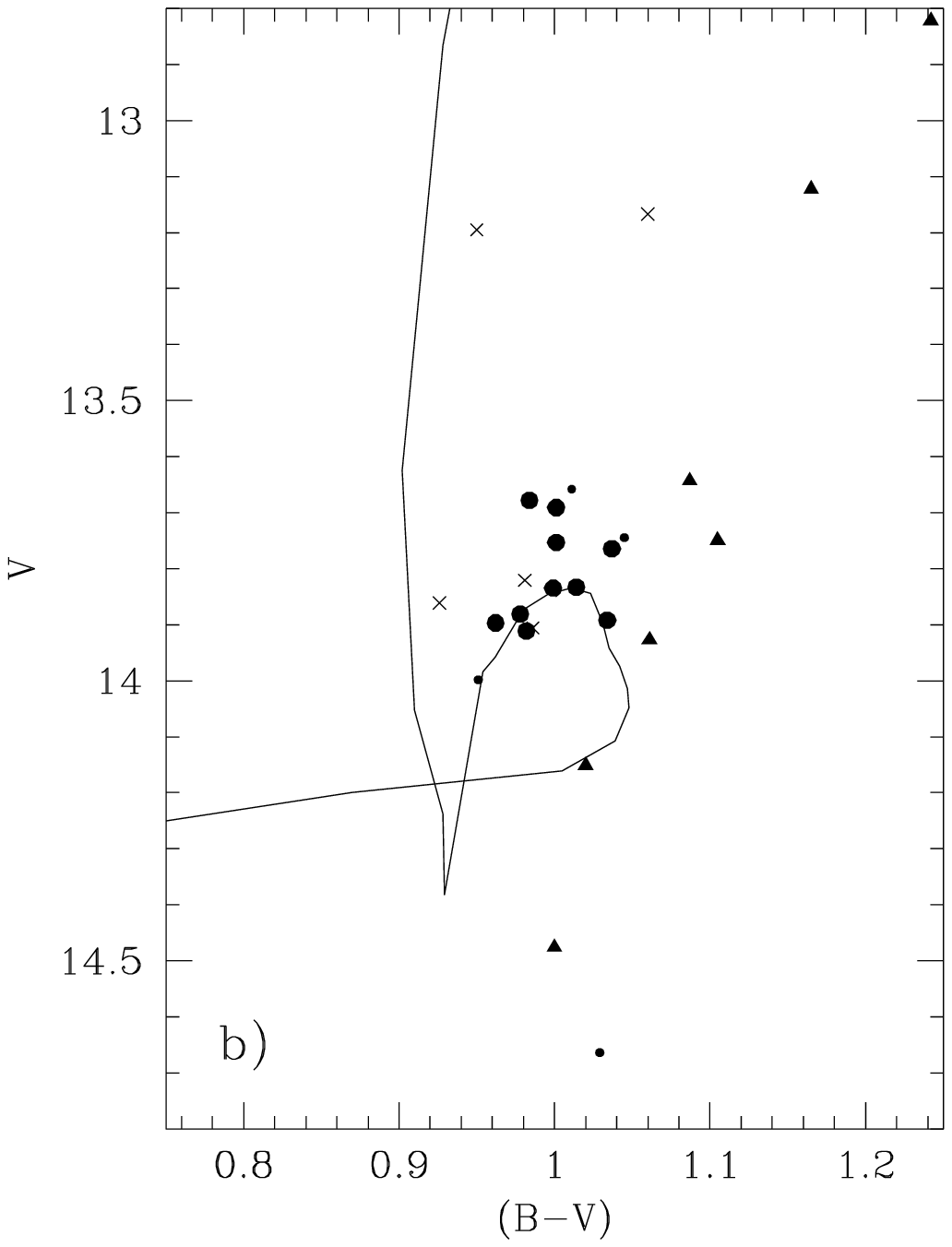}}
\end{minipage} 
\caption{ 
CMDs for the clusters NGC~2660 (a) and NGC~2204 (b). 
Single stars are shown as full dots, known binaries as open simbols.
Triangles denote stars on the ascending red giant branch and 
crosses, non-members according to unpublished CORAVEL radial-velocity 
observations.}
\label{fig_other_clusters}
\end{figure*}

We have looked for other clusters to extend the interpretation of
the clump morphology to further objects. One beautiful example
has been found with NGC~2660, with the CCD photometry of Sandrelli 
et al.\ (\cite{sbtm}). The striking shape of the red giant locus is fairly
well reproduced by the ZAHB for $Z=0.004$ (see 
Fig.~\ref{fig_other_clusters}a). The distribution of the
points leads to masses comprised between 2.2 and $1.9$~\Msun.
IC~1311 also presents a clump with a vertical sequence of stars,
but in absence of membership criterion, it is difficult to separate
the cluster and field stars.

On the oldest side of the age range, NGC~2204 shows a well defined
and compact clump which contains stars on the ZAHB and stars slightly
evolved (Fig.~\ref{fig_other_clusters}b). 

This small sample of representative clusters shows that the
clump morphology changes with ages and that the shape of the ZAHB
predicted by the models, at various chemical compositions, do
reproduce well the observed patterns. Still younger open clusters,
with ages lower than 1~Gyr, have more massive red giants which do not
evolve through the Helium flash and the morphology is again
different. It seems that a single isochrone is also not able to
reproduce the complexity of the clump structure.

\section{Interpretation}
\label{sec_hypothesis}

After suggesting an interpretation to the red clump morphology in
these clusters, it is convenient to further discuss details of the
models. This in order to clarify whether we are really facing 
strange evolutionary behaviours.
 
\begin{figure*}
\resizebox{12cm}{!}{\includegraphics{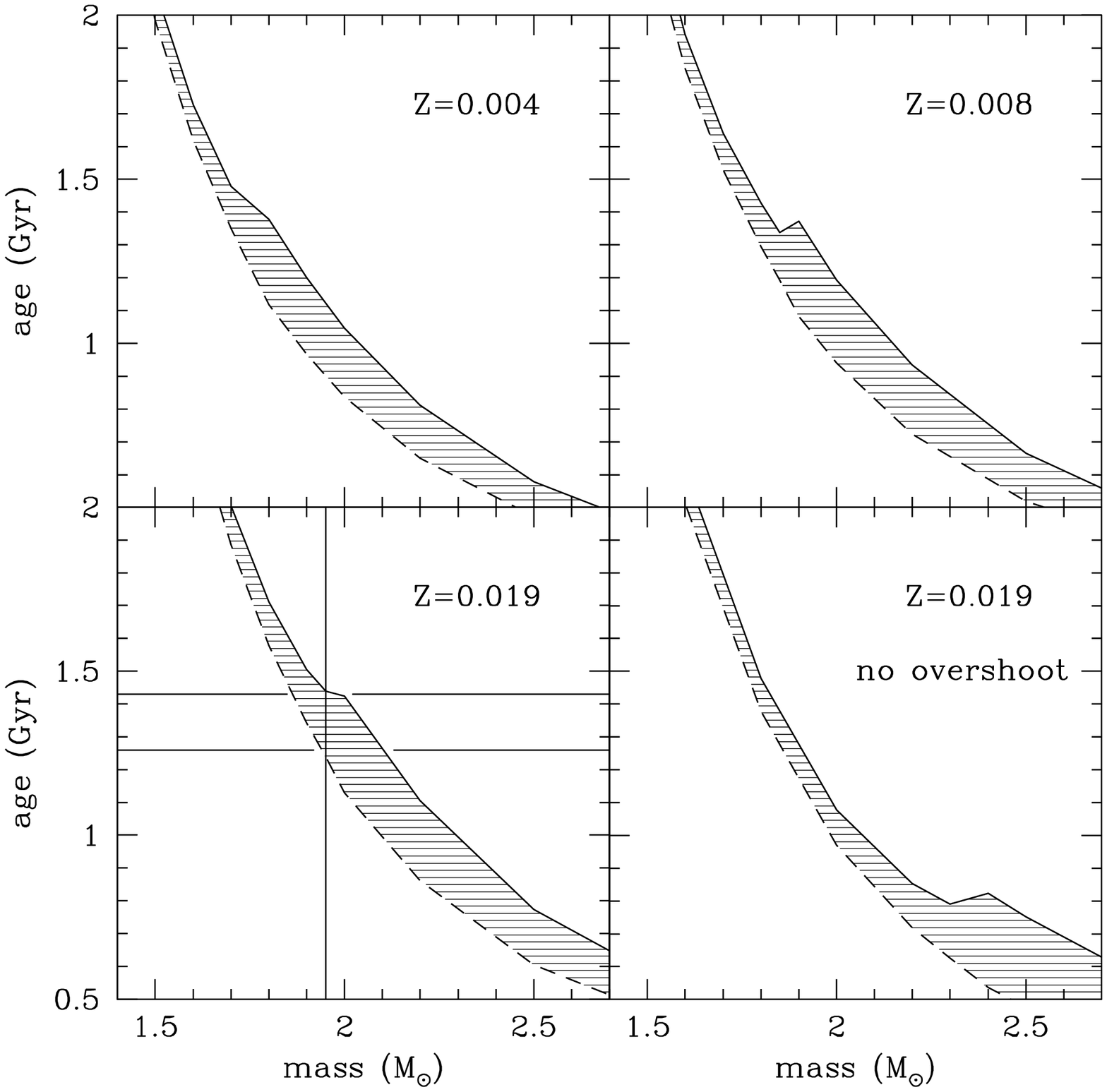}}
\hfill
\parbox[b]{55mm}{
\caption{
The age--mass relation for core-He burning stars (shaded area), 
for 3 values of metallicity: $Z=0.004$, 0.008, and 0.019. The three
first panels are for models which adopt a moderate amount of
convective overshooting. The last 
panel shows the same relation for ``classical'' (i.e.\ without
overshooting) $Z=0.019$ models.
In the panel corresponding to $Z=0.019$ with overshooting, we 
draw a vertical line which schematically separates the low-mass
and intermediate-mass stars. Crossing this line from left to right,
clump stars become less luminous by about 0.4~mag. The two horizontal 
lines limit the age interval in which we can find both types
of clump stars contemporaneously in the same cluster.}
\label{fig_agemass}
}
\end{figure*}

First of all, it is interesting to consider the natural dispersion
of mass in clump stars of different ages. Fig.~\ref{fig_agemass} 
presents the locus
of stars at both the beginning and end of the CHeB stage, on the
age-initial mass diagram. These two lines delimit the region allowed
for clump stars. Singling out a single age for a cluster (i.e.\ a 
horizontal line), we immediately identify the mass range of its clump 
stars. It is interesting to notice that,
at ages lower than about 1~Gyr, this mass range is about 0.2~\Msun\ wide.
When we get to a certain age value (about 1.4~Gyr), it gets suddenly 
narrower, to about 0.1~\Msun. This effect is the 
simple result from the sudden reduction of the CHeB lifetime, by a factor
of about 2.5, which follows the onset of electron-degenerate cores:
this lifetime is of about $10^8$~yr for low mass stars, gets to a maximum
of about $2.5\times10^8$~yr at the transition mass \Mhef, and then
decreases monotonically for stars of higher mass
(see Girardi \& Bertelli~\cite{gb98}; Girardi ~\cite{lg99}). This particular 
behaviour simply reflects the different core masses necessary for igniting 
helium in stars of different masses.   

Fig.~\ref{fig_agemass} then helps to understand the origin of the 
elongated clumps noticed in the first panels of Fig.~\ref{fig_models}: 
they result from the higher dispersion of clump masses found in 
clusters before the transition. 

Of course, there is also an age range in which we find both 
CHeB stars which ignited helium in non-degenerate conditions, and
those which have done it quiescently. This is detailed in the 
lower-left diagram of Fig.~\ref{fig_agemass}. One can notice that,
whereas the main sequence lifetime increases monotonically as we
pass from $\Mhef+\delta M$ to $\Mhef-\delta M$, the CHeB (clump)
lifetime roughly halves as we pass from $\Mhef+\delta M$ to 
$\Mhef-\delta M$. At a given age close to $t(\Mhef)$, we can 
then have the presence
of both clump stars with $\Mhef+\delta M$ at the end of their 
CHeB evolution, 
and stars with $\Mhef-\delta M$ at the beginning of the same 
phase. The interest of this 
situation is that these two kind of stars (with $\Mhef+\delta M$ and
$\Mhef-\delta M$) have CHeB initial 
luminosities differing by up to 0.4~mag, thus providing a good
hint for the origin of dual clumps. The coexistence of both 
CHeB stars lasts for a maximum period of 0.2~Gyr.
Interestingly, in some cases it may happen that stars in both regimes
(degenerate and non-degenerate He ignition), for very short
age intervals, are distributed over 
non-contiguous age intervals. In the Girardi et al.\ (\cite{gbbc}) models, it
happens for the $Z=0.008$ tracks, and for those with 
$Z=0.019$ computed without overshooting (see Fig.~\ref{fig_agemass}), 
for which we have a non-monotonic mass vs.\ age relation
for stars at the late stages of CHeB, in the vicinity of \Mhef.
In the context of the present investigation, this is of course the
most interesting situation, because it could alone 
generate a dual clump {\em in a single isochrone}, without
any artificial assumption about the dispersion of age and mass of
clump stars.

By means of simulations like those shown in Fig.~\ref{fig_models},
however, we have verified that dual clumps do not appear due to 
this effect of mass dispersion, because the clump stars of higher 
mass are found
to be always more evolved than those of lower mass, and hence 
already departed from the ZAHB to much higher luminosities. Thus,
at the age they are observed simultaneously with the brighter clump 
of the lower-mass stars, they no longer represent clump stars of 
lower luminosity.

On the other hand, we should notice that the possibility of having 
dual clumps in single isochrones may depend essentially on 
the velocity at which the clump gets brighter and short-lived
with the stellar mass, or, equivalently, on the velocity at which 
the core mass at the He-flash increases along the transition mass 
\Mhef. The present models (Girardi et al.\ \cite{gbbc})
may not present the level of detail necessary to explore this 
possibility further, due to their limited mass resolution, of 
about 0.05~\Msun. This means that, at timescales 
faster than 0.06~Gyr, our models reflect the result of
interpolating between the contiguous evolutionary tracks, more
than the real evolutionary behaviour of the stars. An improvement 
by a factor of 2 in the model resolution would be desirable.

Other subtle effects can also be invoked
to generate the dual clump features in the models.

\begin{figure*}
\resizebox{\hsize}{!}{\includegraphics{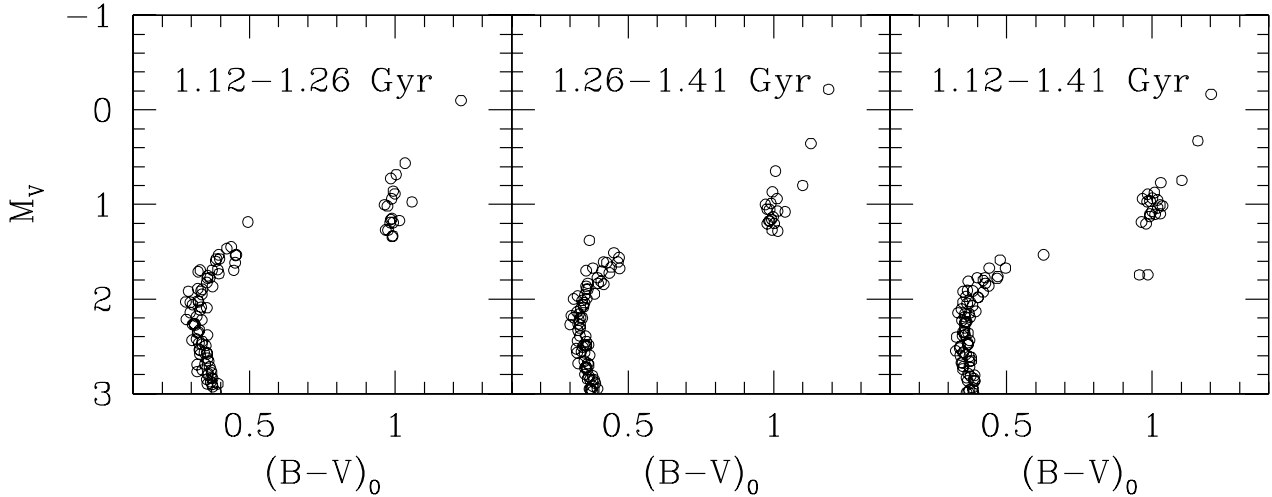}}
\caption{
CMD of models which assume constant star formation between two
age limits: from the left to the right panel, respectively, 
ages are from $1.12$ to $1.26$~Gyr, $1.26$ to $1.41$~Gyr, and 
$1.12$ to $1.41$~Gyr.}
\label{fig_model_deltaage}
\end{figure*}

One of them is the presence of 
a small dispersion of ages for the cluster stars.
It would reflect in a larger range of masses for clump 
stars. Such effect is simulated in Fig.~\ref{fig_model_deltaage},
in which we present synthetic CMDs computed
by assuming constant star formation in the age intervals
$1.12-1.26$, $1.26-1.41$, and $1.12-1.41$~Gyr. These age intervals 
were chosen so that we have the presence of stars in the transition 
region between degenerate and non-degenerate helium ignition.
It can be noticed that the clump is only slightly broadened 
in Fig.~\ref{fig_model_deltaage}, when compared to the simulations 
of Fig.~\ref{fig_models}. Moreover, notice that the assumption of an 
age dispersion implies also that the turn-off region of the CMD is 
slightly broadened, by at most 0.10~mag in colour, if compared to 
the single-age simulations of Fig.~\ref{fig_models}. 

Similar effects, without however any broadening of the MS, can be 
obtained by assuming differential mass-loss by evolved stars.

It is worth remarking that an age spread of $\ga0.1$~Gyr, as assumed
in Fig.~\ref{fig_model_deltaage}, would represent an 
extreme case.  0.01~Gyr would represent a better upper limit to the age 
spread in a cluster, according to estimates based on the pre-main 
sequences of Orion and other very young open clusters (Prosser et al.\ 
\cite{pshs}; Hillenbrand~\cite{lah97}). 

\section{Conclusions}
\label{sec_conclusion}

In this paper, we suggest that the peculiar CMD morphology of the
red clump in the open clusters NGC~752 and NGC~7789, may be indicating
the presence of stars which ignited helium under both degenerate and
non-degenerate conditions. This interpretation is suggested
by the coincidence between the ages of the clusters (as derived 
from the main CMD features), with the ages
at which evolutionary models undergo this main evolutionary
transition. We remark that the event we are referring to, is 
equivalent ot the so-called ``RGB phase transition'' as mentioned
in Renzini \& Buzzoni (\cite{rb86}) and Sweigart et al.\ (\cite{sgr}).

This situation, however, can not be reproduced by models which
assume a single isochrone for the clusters, because the mass 
dispersion of clumps stars in any simple model can hardly be larger
than 0.2~\Msun. Moreover, the tendency found in the simple cluster 
models is that clump stars of higher mass are more evolved away
from the ZAHB, and hence invariably more luminous than stars of lower 
masses. This happens despite that, in the vicinity of the
transition mass \Mhef, more massive clump stars start to burn helium
at luminosities up to 0.4~mag fainter than the less massive ones.

Neither can this situation be simply reproduced assuming an age 
dispersion for the clump stars. The age dispersion required (more 
than 0.1~Gyr) is too large compared to present observational
estimates. Moreover, such a high age dispersion would also cause
a non-negligible --and so far not observed-- spread in the main sequence
region of the CMD.

We are left, then, with a couple of other possibilities. First,
mass-loss on the RGB may cause a significant dispersion of 
clump masses, at a single age. Different amounts of mass-loss
can be triggered, for instance, in stars with different rotational
velocities.  If this is the case, studying clusters like NGC~7789
we may be able to put constraints on the differential mass loss
for stars of this approximate mass range, just in the way we actually
do for stars in globular clusters (cf.\ Renzini \& Fusi Pecci~\cite{rfp}). 
Second, the mass and age at which the transition occurs
is somewhat dependent of the efficiency adopted for overshooting
in stellar cores during the main-sequence phase. Any dispersion in
this efficiency (caused, e.g.\, by different rotational velocities), 
should also reflect on a dispersion of H-exhausted cores masses
at a given age. Such a dispersion would be more evident exactly in
the mass range of the transition, because it is the mass interval in
which the core mass--initial mass relation changes the most. 
For clusters older than 4 Gyr, for instance, the core mass at
He ignition is practically constant and much less sensitive to 
the extension of convective cores.

If any of these interpretations is correct, we face some interesting 
possibilities. First, with more data for open clusters in the 
relevant age range, we may be able to constrain observationally the
velocity at which the transition from non-degenerate to degenerate 
helium ignition occurs. Second, once this age interval is better
documented, we may be able to 
attach independent observables (like the main sequence termination
colour and magnitude) to this transition, to which theoretical
models should comply. The present data for 
NGC~752 and NGC~7789 suggest that $(B-V)_0^{\rm TO}\sim0.5$ at the 
transition mass $\Mhef$, for near-solar metallicities. 
Alternative data for LMC clusters by Corsi et al.\ (\cite{cbfp}), indicates 
$(B-V)_0^{\rm TO}\sim0.25$ for the same transition mass
at a metallicity of about half solar. These numbers seem to be 
reasonably well reproduced by the present models.


\begin{acknowledgements}
We are grateful to Dr A. Bragaglia for providing a copy of the CCD data in
NGC 2660 in advance of publication.
The work by L.G.\ and G.C.\ is funded by the Italian MURST.
L.G.\ acknowledges the hospitality from the Universit\'e de Lausanne
during a visit. The work of J.-C.M. has been supported by grants of the
Swiss National Funds (FNRS).
\end{acknowledgements}


\end{document}